# Non-reciprocal properties of 2D superconductors

Xingrong Ren[1,2], Huiqing Ye[3,4,†], Tian Le[1,2,†]

[1]*Center for Quantum Matter and School of Physics, Zhejiang University, Hangzhou 310058, China*
[2]*Institute for Advanced Study in Physics, Zhejiang University, Hangzhou 310027, China*
[3]*Center for Correlated Matter and School of Physics, Zhejiang University, Hangzhou 310058, China*
[4]*School of Physics, Hangzhou Normal University, 310036, Hangzhou, China*

**Two-dimensional (2D) superconductors, characterized by their inherent quantum confinement, strong spin-orbit coupling, and diverse forms of symmetry breaking, provide an ideal platform for exploring novel quantum transport phenomena. This review summarizes recent experimental progress in the non-reciprocal properties of 2D superconductors, focusing on second harmonic resistance in the resistive superconducting state and the supercurrent diode effect (SDE) in the dissipationless superconducting regime. We discuss the various origins of these phenomena, distinguishing between intrinsic mechanisms, such as finite-momentum Cooper pairing, and extrinsic mechanisms driven by asymmetric vortex dynamics and device geometry. We present a systematic classification of zero-field SDE into polarity-reversed and polarity-locked behaviors, a distinction governed by the interplay between intrinsic time-reversal symmetry breaking and external magnetic response. Furthermore, we examine how the efficiency and polarity of SDE are modulated by tuning parameters including magnetic/electric fields, strain, device geometry, thermodynamic conditions, and microwave irradiation. We conclude by highlighting the application potential of these tunable diodes in high-efficiency rectification, superconducting logic, and neuromorphic computing.**

## 1. Introduction

Non-reciprocal transport is a fundamental phenomenon where the current-voltage (*I-V*) characteristics depend on the direction of the bias current or voltage, meaning $I(+V) \neq I(-V)$ or $V(+I) \neq V(-I)$. In general, a system must break inversion symmetry (IS) to achieve non-reciprocity. A prime example of this is the voltage-controlled p-n junction, where non-reciprocity arises because the depletion layer width changes differently in response to a positive voltage ($V^+$) versus a negative voltage ($V^-$).[1] In this case, the voltage-controlled conduction process preserves time-reversal symmetry (TRS). However, the Onsager relations dictate that current-controlled non-reciprocity requires breaking both IS and TRS.[2, 3] This phenomenon is widely observed in non-centrosymmetric quantum materials under a magnetic field *B*,[4] including silicon field-effect transistors,[5] magnetic bilayers,[6] polar semiconductors,[7, 8] topological insulators[9] and chiral crystals.[10-13] In these systems, non-reciprocal behavior manifests as magnetochiral anisotropy (MCA)[10, 14] or unidirectional magnetoresistance[10] as illustrated in Fig. 1(a), driven mainly by the spin-orbit coupling (SOC) and magnetic energy scales relative to the Fermi energy $\varepsilon_F$.

---

[†] Corresponding author. E-mail: huiqingye@zju.edu.cn
[†] Corresponding author. E-mail: letian_phy@zju.edu.cn

Superconductivity introduces the superconducting (SC) gap $\Delta$, an energy scale much smaller than $\varepsilon_F$. As shown in Fig. 1(b), this significantly smaller energy scale results in a dramatic enhancement of the non-reciprocal signal within the resistive SC state near the transition temperature, $T_c$.[15] Recent years have witnessed a surge of interest in non-reciprocal transport in SC systems, a field marked by the emergence of striking phenomena like the second harmonic resistance (SHR) [4] and the supercurrent diode effect (SDE) (Fig. 1(c)).[16, 17] Observing a pronounced non-reciprocal signal requires high current densities with minimal heating effect. Two-dimensional (2D) superconductors are an ideal platform, as their reduced thickness intrinsically satisfies these requirements. In particular, 2D superconductors host a wealth of exotic electronic properties that are pivotal for realizing non-reciprocal transport, most notably strong SOC and spontaneous symmetry breaking.[18] The category of "2D superconductors" comprises a variety of systems, including, but not limited to, thin films, nanoplates and heterostructure interfaces.

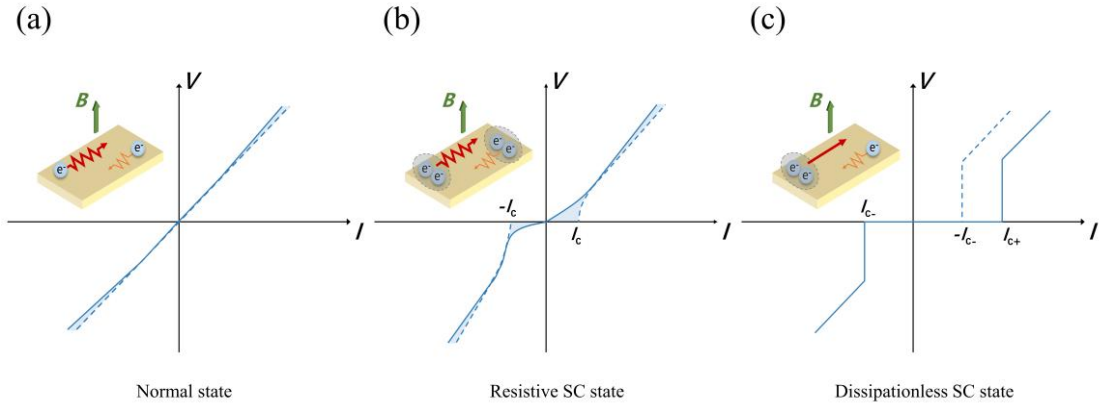

Fig. 1 (a) Non-reciprocal *V-I* curve in the normal state. The dashed line indicates the linear response. Inset: Schematic of non-reciprocal electron motion. (b) Non-reciprocal *V-I* curve near $T_c$, where the vortex matter produces a resistive SC state. The dashed line denotes the corresponding reciprocal (symmetric) response for reference. Inset: Schematic of non-reciprocal Cooper pair motion. (c) Non-reciprocal *V-I* curve in the zero-resistance state, demonstrating the SDE. The dashed line represents the typical V-I curve in the absence of SDE. Inset: Schematic of unidirectional supercurrent.

This review summarizes recent experimental progress on non-reciprocal phenomena in 2D superconductors, with a focus on second harmonic resistance (SHR) and the supercurrent diode effect (SDE). The reported SHR is discussed in both heterostructure-free systems and in superconducting heterostructures. Furthermore, the SDE is reviewed across a wide range of systems, covering field-assisted and zero-field mechanisms. Notably, in systems exhibiting a zero-field SDE, the response under an applied magnetic field can be classified as polarity-reversed and polarity-locked behaviors. The discussion extends to how the efficiency and polarity of the SDE are modulated by parameters such as magnetic/electric fields, strain, device geometry, thermodynamic conditions and microwave irradiation. Finally, we discuss the broad tunability of the SDE, which highlights its potential for future diverse applications.

## 2. Second-harmonic resistance

In the *I-V* characteristics of 2D superconductors, non-reciprocity gives rise to a second-harmonic resistance (SHR), a phenomenon intimately tied to vortex matter under IS breaking.[4, 15, 19] Importantly, IS breaking is a necessary condition for SHR, whereas TRS not necessarily need to be broken. This distinction stems from the fact that SHR originates from a directional asymmetry in vortex motion within an IS-broken environment, even in the absence of external magnetic fields or intrinsic magnetic order. Experimentally, SHR is defined as $R^{2\omega} \equiv \frac{V^{2\omega}}{I_0}$, where $V^{2\omega}$ represents the second-harmonic voltage component measured via lock-in techniques, and $I_0$ denotes the amplitude of the applied AC current. The vortex matter encompasses unpinned vortices driven by an external magnetic field, as well as thermally activated vortex-antivortex pairs near $T_c$. Both types of vortex motion can create a resistive state and produce a detectable SHR signal. The underlying IS breaking in such non-reciprocal systems stems from two distinct sources: intrinsic non-centrosymmetric crystal structures and engineered asymmetric interfaces. These aspects will be elaborated on in following sections.

### 2.1. Heterostructure-free superconductors

In an intrinsic non-centrosymmetric superconductor, IS is broken by the inherent crystal structure or by the electronic states. A representative non-reciprocal signal is the enhanced SHR, which was first observed in the resistive SC fluctuation regime of electrostatically doped $MoS_2$.[15] In an electric double-layer transistor (EDLT) configuration, a multilayer $MoS_2$ flake with the 2H polytype can be gated into a SC state, as shown in Fig. 2(a) and (b). The applied electric field breaks out-of-plane inversion symmetry, rendering adjacent monolayers inequivalent. Consequently, the magnetic field causes $R^{2\omega}$ to initially rise, reach a maximum at a finite *B*, and finally decrease, as illustrated in Fig. 2(c). The magnitude of non-reciprocity is quantified by the parameter γ, defined as γ=$(2R^{2\omega})/(R^{\omega} BI_0)$. In the SC state, γ reaches values on the order of $10^3$ $T^{-1}$ $A^{-1}$, with a maximum close to 8000 $T^{-1}$ $A^{-1}$, approximately five orders of magnitude larger than those in non-SC materials. The SHR in $MoS_2$ is dominated by the ratchet motion of vortices, that arises from an asymmetric pinning potential.[20]

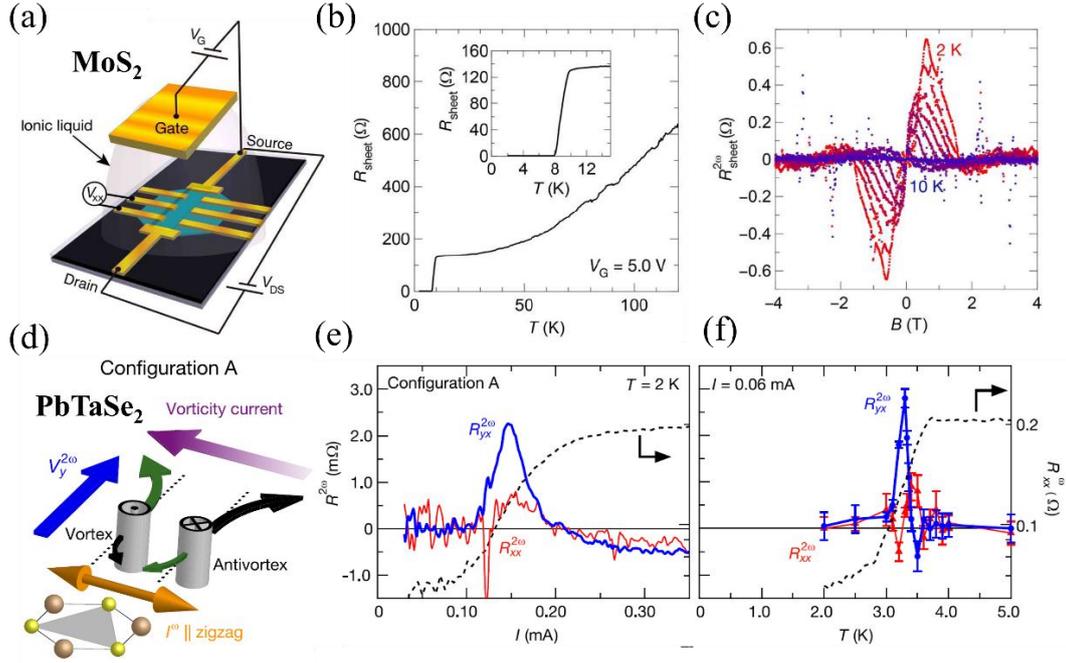

Fig. 2 (a) Schematic image of a MoS$_2$-EDLT. (b) Temperature dependence of the resistance for a MoS$_2$-EDLT at a gate voltage of 5.0 V. Inset: A detailed view of the SC transition. (c) Antisymmetrized second harmonic magnetoresistance $R^{2\omega}$ at 2 to 10 K. Reproduced with permission from Ref. [15]. Copyright 2017, The Authors. (d) Schematic illustration of the asymmetric Hall effect of vortex-antivortex string pairs in configuration A. (e) Current dependence of longitudinal and transverse $R^{2\omega}$ at $T$ = 2 K in a configuration A. (f) Temperature dependence of longitudinal and transverse $R^{2\omega}$ in configuration A. Reproduced with permission from Ref. [21]. Copyright 2022, The Authors.

Similar behavior has been identified in various superconductors, such as PbTaSe$_2$,[22] NbSe$_2$,[23] WS$_2$,[24] SrTiO$_3$,[25] and CsV$_3$Sb$_5$.[26] These results confirm that SHR is intimately related to vortex motion under an applied $B$, suggesting that non-reciprocal transport is a powerful tool for investigating vortex dynamics. More intriguingly, SHR has also been observed under TRS in PbTaSe$_2$, as revealed by Yuki *et al.* near the critical supercurrent $I_c$ and $T_c$ (Figs. 2(e) and 2(f)).[21] This phenomenon is attributed to the asymmetric Hall effect of vortex-antivortex string pairs in the resistive SC state, as illustrated in Fig. 2(d). Strikingly, a recent study on 2H-Ta$_2$S$_3$Se further demonstrates zero-field nonreciprocal transport in an anomalous metallic state, underscoring the deep relation between spontaneous TRS and vortex dynamics even in the absence of global superconductivity and external magnetic field.[27]

## 2.2. Superconducting heterostructures

IS breaking naturally occurs at the interface of two different materials, originating from their internal atomic asymmetry. In several topological insulator (TI)/superconductor heterostructures, such as Bi$_2$Te$_3$/FeTe,[28] (Bi$_{1-x}$Sb$_x$)$_2$Te$_3$/FeSe$_{0.1}$Te$_{0.9}$,[29] and Bi$_2$Te$_3$/PdTe$_2$,[30] a giant non-reciprocal signal appears in the resistive SC state. Crucially, this non-reciprocity stems from the intrinsic electronic asymmetry of the topological surface states, specifically, the spin-momentum locking works together with SC fluctuations. This interpretation is strongly supported by the marked dependence of the non-reciprocal coefficient on the TI layer thickness and the reversal of its sign when the Fermi level crosses the Dirac point, both of which underscore the essential role of the topological surface state. These findings demonstrate a direct connection between superconductivity and the topological nature of the electronic states.

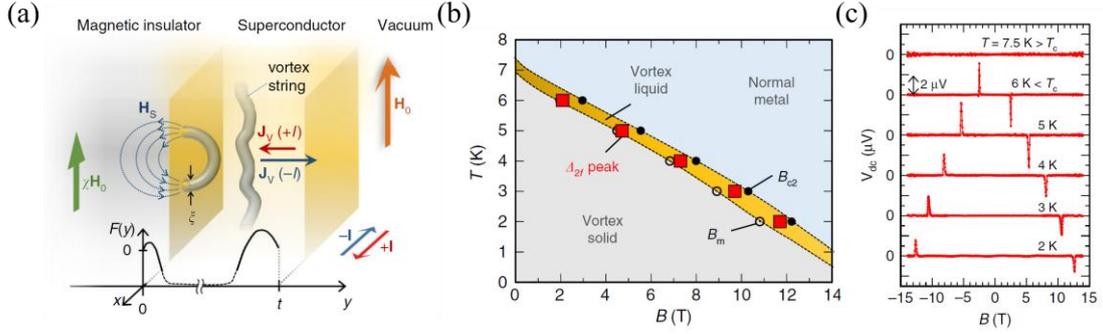

Fig. 3(a) Non-reciprocity in an asymmetric external environment. A superconductor carrying vortex strings is placed adjacent to a magnetic insulator (susceptibility $\chi$) under an external field $H_0$. The difference in vortex nucleation energy $F$ at the two interfaces leads to a non-reciprocity vortex flow $J_V$ when driven by a current $\pm I$. (b) The $B$–$T$ phase diagram of MoGe layer and the appearance of non-reciprocal resistance, represented by the $\Delta_{2f}$ peak. (c) Magnetic field dependence of the voltage generated in the MoGe layer at selected temperatures. Reproduced with permission from Ref. [31]. Copyright 2018, The Authors.

Remarkably, non-reciprocity can even arise in heterostructures based on amorphous films, as exemplified by the MoGe/$Y_3Fe_5O_{12}$.[31] In this system, the asymmetric magnetic environment across the SC film creates a difference in the vortex nucleation energy at the interface, as illustrated in Fig. 3(a). This asymmetry not only enables SHR but also generates a DC voltage through the rectification of ambient electromagnetic fluctuations (Fig. 3(b) and (c)). This work establishes that a global symmetry-breaking environment is sufficient to induce pronounced non-reciprocal effects.

## 3. Supercurrent diode effect

The previously discussed SHR of 2D superconductors is a hallmark of non-reciprocity in the dissipative regime. Recently, non-reciprocal superconductivity has extended to the dissipationless zero-resistance state, where it manifests as a directional dependence of $I_c$, known as the SDE. In this effect, the critical supercurrent $I_c$ for one direction ($I_c^+$) is different from that in the opposite direction ($I_c^-$), establishing a one-way street for dissipationless flow, as illustrated in Fig. 1(c). The conceptual foundation for a supercurrent diode was laid in 2007 by Hu *et al.*, who theorized a diode effect based on p-type and n-type Josephson junctions (JJs).[32] However, the definitive experimental demonstration was achieved in 2020 by Ando *et al.*, using a noncentrosymmetric [Nb/V/Ta]$_n$ superlattice, sparking intensive research into this non-reciprocal phenomenon across diverse 2D SC materials.[33]

## 3.1. Inversion symmetry breaking and external magnetic field

The SDE emerges from the simultaneous breaking of TRS and IS in the dissipationless SC state. An external magnetic field usually provides the most straightforward way to break TRS, whereas the necessary IS breaking can be realized either through intrinsic material properties or by extrinsic engineering. Intrinsic SDE is directly tied to the fundamental nature of the SC order parameter, such as the formation of finite-momentum Cooper pairs (FMCP). In contrast, the extrinsic type is governed by the interaction of the supercurrent with an asymmetric potential environment within the device.

### 3.1.1. Intrinsic mechanisms

*Finite-momentum Cooper pairs:* In superconductors with Rashba-type SOC, an in-plane magnetic field lifts Kramers degeneracy and displaces spin-split Fermi surfaces in momentum space, resulting in Cooper pairs with a finite centre-of-mass momentum $q_0$, analogous to the Fulde-Ferrell-Larkin-Ovchinnikov (FFLO) state.[34-36] In Ising superconductors such as monolayer NbSe$_2$, the strong Ising-type SOC pins electron spins out-of-plane. The generation of FMCP requires the combination of an out-of-plane magnetic exchange field (typically induced via ferromagnetic proximity) and an in-plane magnetic field.[37-40] JJs incorporating centrosymmetric topological semimetals can also support such pairing via Zeeman effects or Meissner screening currents acting on spin-momentum-locked surface states, without requiring bulk inversion asymmetry.[39, 41-45]

Across these systems, the depairing energy depends directionally on $q_0$, yielding a non-reciprocal $I_c$.[39, 43, 46, 47] A key signature of this intrinsic mechanism is its non-monotonic $B$-dependence, including polarity reversals at finite fields.[46] The SDE associated with this mechanism could exist in noncentrosymmetric superlattices (Fig. 4(a)),[33] Ising superconductors (Fig. 4(b)),[40, 48] and topological material-based JJs (Fig. 4(c)).[49]

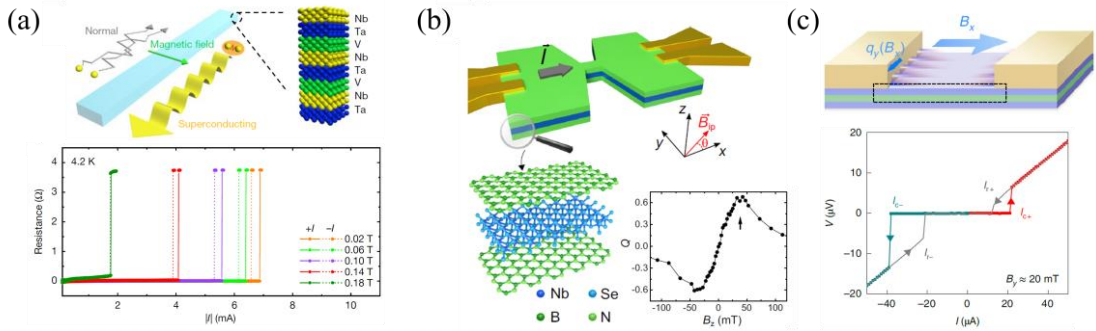

Fig. 4(a) Top panel: Schematic of the Rashba SOC-controlled SDE. The artificial [Nb/V/Ta]$_n$ superlattice exhibits global IS breaking along the stacking direction. A unidirectional flow of Cooper pairs emerges when the current, magnetic field, and symmetry-breaking axis are mutually orthogonal. Bottom panel: Resistance as a function of positive and negative current under various magnetic fields at 4.2 K. Reproduced with permission from Ref. [33]. Copyright 2020, Springer Nature. (b) Schematic of the NbSe$_2$ device, comprising a stack of hBN/NbSe$_2$/hBN. The NbSe$_2$ flake

has a thickness of 2, 3, or 5 layers. Lower-right inset: The SDE efficiency as a function of $B_z$ reaches a maximum value of ~60%. Reproduced with permission from Ref. [48]. Copyright 2022, the authors. (c) Top panel: Schematic of the JJ under an in-plane $B$ ($B||I$), producing a FMCP along the $y$-direction. Bottom panel: $I$-$V$ curve of a topological NiTe$_2$-based JJ device, showing a large non-reciprocal supercurrent under an in-plane $B$. Reproduced with permission from Ref. [49]. Copyright 2022, the authors.

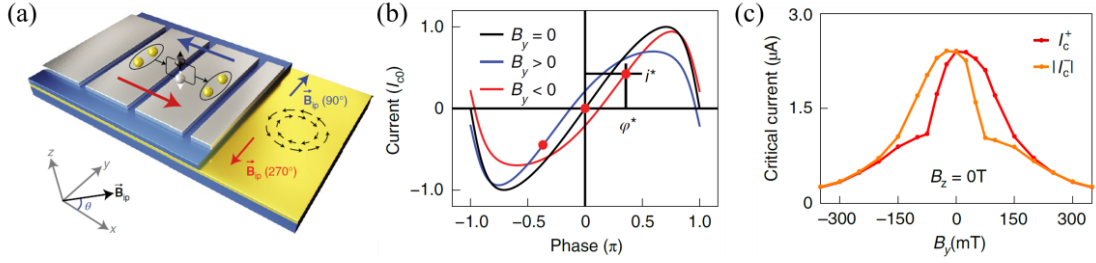

Fig. 5(a) Schematic of a JJ array consisting of a chain of grey Al islands on a yellow InAs quantum well. The competition between an in-plane magnetic field $B_{ip}$ and the inherent Rashba spin-texture (indicated by counterpropagating circles of black arrows) governs the strength and direction of the spontaneous supercurrents, represented by red and blue arrows. (b) Illustrative CPR for a JJ with high transparency and strong SOC. The black curve represents the symmetric CPR at zero $B$, while the red and blue curves show the skewed CPR under in-plane $B$ of opposite directions $B_y||y$ ($B_y > 0$ and $B_y < 0$, respectively). (c) $I_c^+$ and $I_c^-$ as a function of $B_y$ for $B_z = 0$. Reproduced with permission from Ref. [50]. Copyright 2022, the authors.

*Anomalous current-phase relations (CPR):* In JJs, the SDE originates from fundamental modifications to the CPR.[51] When SOC and TRS breaking coexist, the conventional sinusoidal CPR develops two key characteristics: significant higher-harmonic components (particularly a $\sin(2\phi)$ term) and an anomalous phase shift $\phi_0$.[52-56] The interplay between these terms breaks the antisymmetric nature of the CPR, resulting in $I(\phi) \neq -I(-\phi)$ and consequently $I_c^+ \neq I_c^-$, as typically shown in Fig. 5.[50, 57-59]

This mechanism is particularly pronounced in two types of JJs: highly transparent interfaces where higher harmonics occur naturally, and junctions containing topological materials or strong SOC barriers. Representative systems demonstrating this physics include JJs based on semiconductors[50, 60-63] and topological semimetals.[64-67] Furthermore, in topological JJs, Majorana bound states can dramatically enhance the $\sin(2\phi)$ component, leading to exceptionally high diode efficiency.[68-71]

### 3.1.2. Extrinsic mechanisms

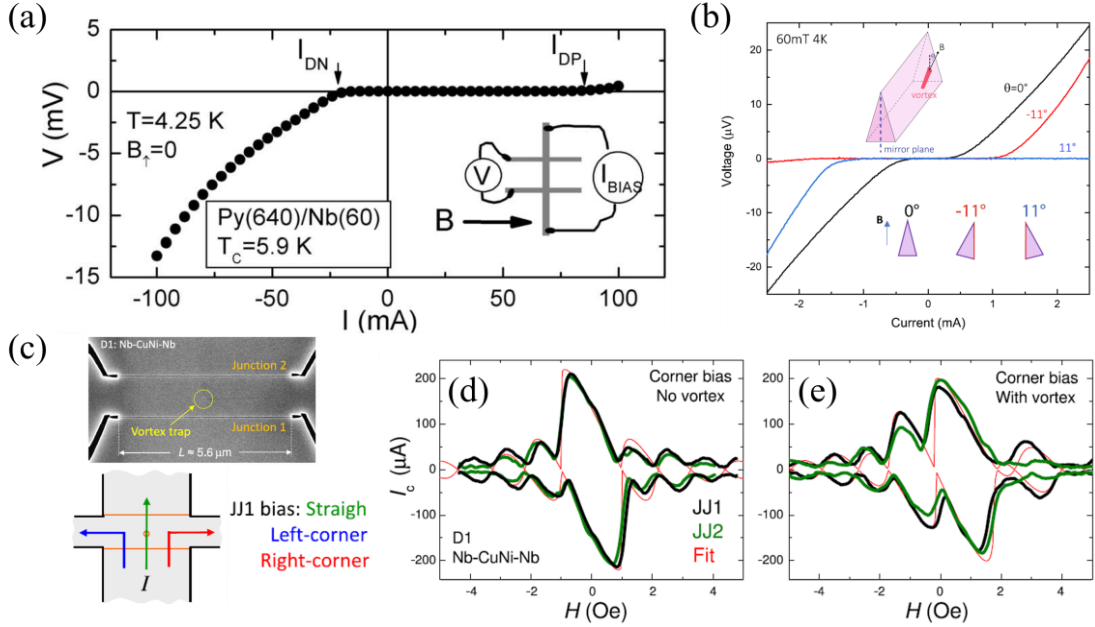

Fig. 6(a) Asymmetric *V-I* curve of a Py/Nb bilayer at 4.25 K, revealing a large disparity between positive and negative depinning currents ($I_{DN}$ and $I_{DP}$). The geometry of the bilayer is shown in the inset. Reproduced with permission from Ref. [72]. Copyright 2009, American Institute of Physics. (b) *V-I* curves demonstrating the SDE. A pronounced non-reciprocity appears at $\theta = 11°$, showing vastly different $I_c$ for opposite current polarities. Inset: Schematic of the SC wire with a precisely engineered triangular cross-section, made of amorphous W-C-O ($T_c \approx 5$ K). The magnetic field angle $\theta$ is defined with respect to the wire's mirror plane. Reproduced with permission from Ref.[73]. Copyright 2025, the authors. (c) Schematic of the cross-shaped device: top panel shows an SEM image of the Nb-CuNi-Nb junctions with a central vortex trap; bottom panel illustrates the bias configurations for controlling the current-induced self-field. (d) and (e) Modulation of the SDE, manifested as asymmetric $I_c(H)$ patterns, under right-terminal bias (d) without and (e) with a vortex trapped in the center. Data for junctions 1 and 2 are shown in black and olive, respectively. Red lines are numerical fits. Reproduced with permission from Ref. [74]. Copyright 2022, the authors.

*Asymmetric vortex dynamics:* In type-II superconductors under out-of-plane *B*, the dissipative state emerges when magnetic vortices begin to move.[75] By introducing artificially engineered asymmetric pinning centers, a directional potential gradient is created. This gradient causes different $I_c$ for vortex depinning along opposite directions, a phenomenon known as the "ratchet effect".[76-79] This mechanism drives the SDE, also referred to as the vortex diode effect, which was originally proposed by Edelstein in the 1990s.[72, 80-85] Such asymmetric pinning can be realized through conformal mapped nanohole arrays,[86, 87] asymmetric geometries,[73, 88, 89] or integrated nanostructures.[90-92] These approaches have been demonstrated in conventional SC films[93-99] and superconductor/ferromagnet hybrids,[88, 100-104] where rectification efficiency is widely tunable through nanofabrication (Figs. 6(a) and 6(b)).

*Asymmetric Josephson junctions:* Extrinsic SDE in JJs typically stems from spatially inhomogeneous magnetic environments or geometric asymmetry. For instance, in planar junctions with trapped Abrikosov vortices, stray fields and asymmetric bias currents break inversion symmetry (Figs. 6(c), 6(d) and 6(e)).[74, 105, 106] Alternatively, non-reciprocity can arise from spin filtering in magnetic Josephson junctions[107-109] or asymmetric quasiparticle current in atomic scale junctions.[110] These mechanisms generate non-reciprocal $I_c$ without requiring SOC or anomalous CPR.

### 3.1.3. Distinguishing intrinsic and extrinsic mechanisms in experiments

The experimental observation of the SDE under a magnetic field raises a pivotal question regarding its origin: does it arise from an intrinsic property of the superconducting state (e.g., finite-momentum pairing, anomalous CPR) or from extrinsic, device asymmetries (e.g., geometry, vortex pinning, edge barriers)? Disentangling these contributions is challenging but crucial for validating novel physical mechanisms.

In general, intrinsic mechanisms should be largely independent of macroscopic device geometry (e.g., width, length, contact placement). For example, the SDE arising from FMCP typically depends on the vector relationship between the applied $B$, the crystallographic axes, and the current direction.[42, 49] Such an intrinsic SDE can exhibit a non-monotonic $B$-dependence and may reverse polarity at certain field value.[46] In non-centrosymmetric crystals, the intrinsic SDE must respect the point group symmetry. Rotating the current direction relative to the crystal axes should modulate the diode response.[111] In 2D materials, intrinsic mechanisms (e.g., Ising SOC) should show a pronounced evolution with layer number.[37] Direct measurement of the CPR in JJs can reveal intrinsic anomalous phase shifts or higher harmonics, providing a crucial signature for intrinsic diode mechanisms.[50]

By contrary, extrinsic SDE driven by vortex pinning or geometric asymmetry typically displays device dependent behaviour. They are highly sensitive to device layout, edge quality and the detailed configuration of pinning centers. The SDE polarity and magnitude can vary dramatically between nominally identical devices, and may even change after current training, reflecting the stochastic nature of vortex trapping.[73, 74, 93]

### 3.2. Dual broken inversion and time-reversal symmetry at $B$ = 0 T

External magnetic fields provide the most direct approach to breaking TRS, yet achieving the SDE at zero magnetic field is essential for practical applications. This requires materials or heterostructures that intrinsically possess or can be engineered to exhibit simultaneous breaking of IS and TRS. Zero-field SDE can be generally classified into two categories based on the relationship between their intrinsic TRS breaking and the response to an external $B$. This distinction clarifies the underlying physical origins and guides device design.

### 3.2.1. Polarity-reversed behaviors

In this category, internal TRS breaking originates from ferromagnetic moments or unconventional SC order parameters. It is important to note that vortex pinning can also contribute to a zero-field SDE, typically observed after magnetic field sweeps or current pulses. This extrinsic contribution must be examined carefully.[112-114] For systems containing magnetic moments, the SDE polarity is governed by the magnetization direction and can be reversibly switched by the training field.[40, 100, 115-119] A prominent example is the [Nb/V/Co/V/Ta]$_n$ multilayer, where the ferromagnetic Co layers break TRS, creating a non-volatile, switchable diode (Figs. 7(a), 7(b) and 7(c)).[100] Magnetic proximity effects also enable electrically programmable zero-field SDE in heterostructures like NbSe$_2$/Fe$_3$GeTe$_2$, where current-induced spin-orbit torques flip the magnetization.[40] Notably, twisted bilayer materials have also been shown to host spontaneous spin polarization and collective ferromagnetism, providing a twist-induced zero-field SDE without the need for extrinsic magnetic layers (Figs. 7(d) and 7(e)). [120-123] Additionally, in Fe(Te, Se)-based JJs, the zero-field SDE arises from the interplay between the interfacial ferromagnetism and superconductivity. In this case, the stochastic polarity provides direct evidence of the spontaneous TRS breaking at the interface.[118]

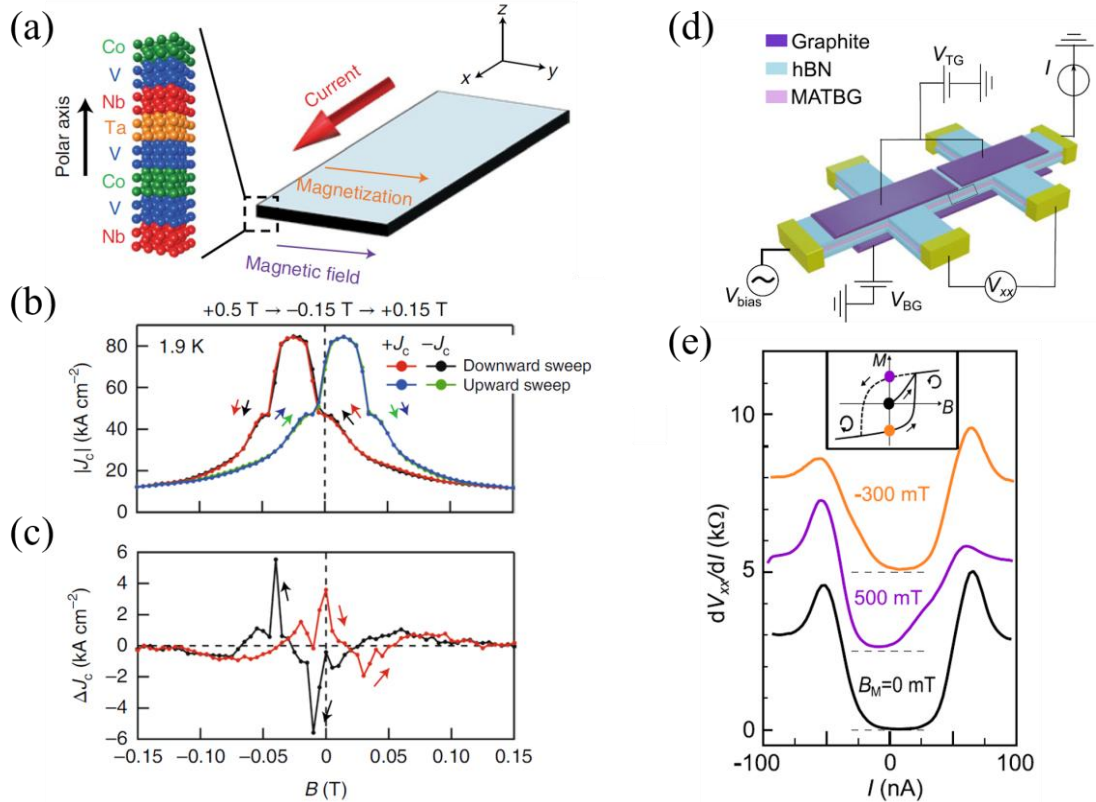

Fig. 7 (a) Schematic of the measurement configuration for SDE. The magnetic field is applied perpendicular to both the current and the polar axis. (b) Magnetic field dependences of critical current density $J_c$ for positive and negative currents at 1.9 K. (c) Magnetic field dependence of $\Delta J_c$, derived from Fig. 7(b). Reproduced with permission from Ref. [100]. Copyright 2022, the authors.
(d) Schematic of the device and measurement circuit. (e) $dV_{xx}/dI$ vs. $I$ at $B = 0$ mT and $T = 800$ mK, measured after cooldown (black) and after exposure to two opposing pre-magnetizing fields $B_M$. The curves are vertically shifted by 2.5 kΩ for clarity. Inset: Magnetization versus $B$, with colored dots marking the magnetic states during measurement and arrows indicating the field-sweep direction. Reproduced with permission from Ref. [123]. Copyright 2023, the authors.

Apart from magnetic systems, intrinsic TRS breaking in unconventional superconductors provides another route to zero-field SDE.[122-126] In small-twist-angle trilayer graphene, correlated states with spontaneous TRS breaking support switchable diode effects whose polarity can be controlled by field training.[122] Similarly, in cuprate systems like twisted $Bi_2Sr_2CaCu_2O_{8+\delta}$ (BSCCO), JJs exhibit non-reciprocal transport signatures of TRS breaking.[124] The kagome superconductor $CsV_3Sb_5$ also shows zero-field SDE modulated by thermal cycling, attributed to reorganization of chiral SC domains with spontaneous TRS breaking.[125] These materials demonstrate that intrinsic SC order parameters can generate and control non-reciprocal supercurrents in the absence of external magnetic fields.

### 3.2.2. Polarity-locked behaviors

This category includes zero-field SDE characterized by $B$-even behavior, where diode polarity remains fixed and does not switch with the magnetic field direction. Key material systems demonstrating this behavior include $Nb_3Br_8$-based JJs,[127] $NiI_2$-based JJs,[128] strain-engineered $PbTaSe_2$,[111] $NbSe_2$ devices,[129-131] cuprates flakes,[132] Iron-based SC devices,[133, 134] and Cooper-pair transistors.[135] The underlying physical mechanisms are diverse, encompassing the following:

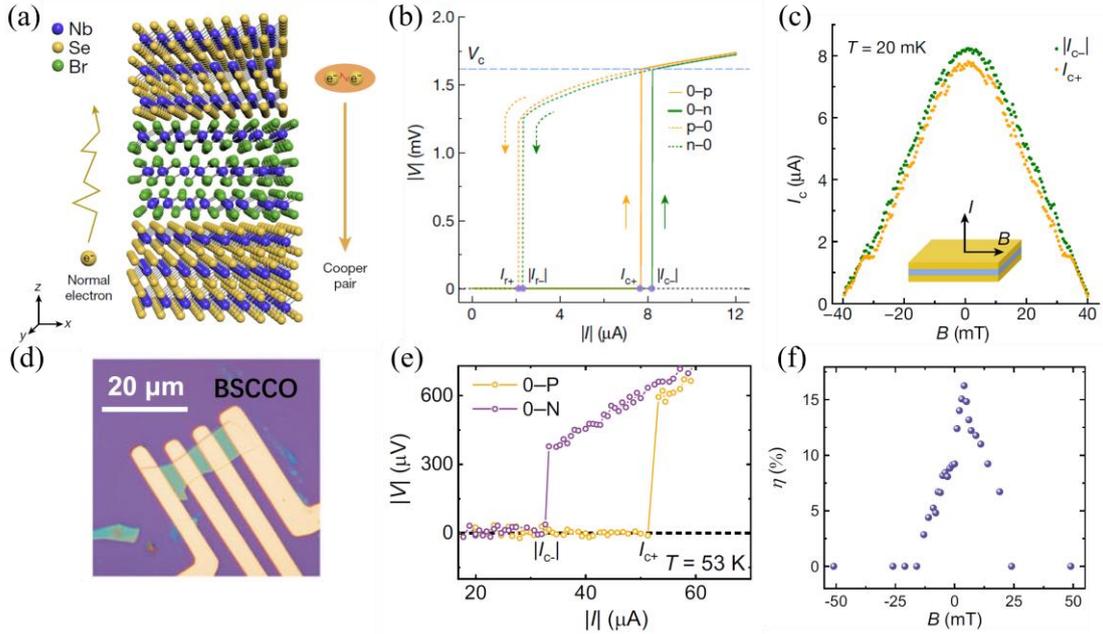

Fig. 8 (a) Schematic of a $NbSe_2/Nb_3Br_8/NbSe_2$ JJ. (b) Non-reciprocal $V$–$I$ curves at zero field, showing distinct $I_c$ for positive and negative sweep directions. (c) $B$ dependence of $I_{c+}$ (orange dots) and $|I_{c-}|$ (green dots), revealing a $B$-even SDE behavior. Inset: measurement geometry. The orange and blue layers represent $NbSe_2$ and $Nb_3Br_8$, respectively. Reproduced with permission from Ref. [127]. Copyright 2022, the authors. (d) Optical microscope image of the BSCCO device on $SiO_2$/Si substrate. (e) Non-reciprocal $V$–$I$ curves at zero field, demonstrating the SDE. (f) Diode efficiency $\eta$ as a function of perpendicular $B$. Reproduced with permission from Ref. [132]. Copyright 2025, the authors.

**Voltage-controlled Mechanisms:** Voltage-controlled non-reciprocal conduction processes (i.e., $I(+V) \neq I(-V)$) typically preserve TRS,[15, 136, 137] as seen in theoretical proposals for p-n-type JJs based on cuprates.[32] Within this framework, a zero-field SDE can arise from a self-organized Mott-insulating depletion region formed at the interface between two superconductors.

**Electric polarization:** Electric polarization induced by strain or structural asymmetry establishes a built-in electric field along specific crystallographic directions, which acts as an effective internal bias voltage. This field breaks inversion symmetry locally and asymmetrically modifies superconducting transport properties such as the critical supercurrent density and Josephson coupling, without requiring an external magnetic field. (Figs. 8(a), 8(b) and 8(c)).[111, 127, 129-131, 134, 136]

**Shift Current:** The shift current, arising from the Berry phase of the electronic states in noncentrosymmetric crystals, can generate a non-reciprocal supercurrent response. It contributes directly to the difference between $I_{c+}$ and $I_{c-}$ even without an external magnetic field. This quantum geometric effect produces a fixed SDE polarity that is fundamentally determined by the electronic structure.[17, 138-141]

**Insensitive magnetic moments:** Magnetic moments can be strongly pinned by the underlying charge or orbital order, making them insensitive to external $B$. This robust magnetic ordering provides a mechanism for spontaneous TRS breaking that is locked to the crystal lattice, thereby generating a $B$-even zero-field SDE. Compelling evidence for this scenario comes from BSCCO, where a fixed-polarity zero-field SDE has been interpreted as a signature of an intrinsic loop-current order (Figs. 8(d), 8(e) and 8(f)).[132] A similar behavior is observed in chiral molecule-intercalated $TaS_2$ hybrid superlattices. In these systems, structural chirality induces an unconventional SC state with intrinsic TRS breaking.[126]

**Extrinsic mechanisms:** Circuit-level effects can also create a zero-field SDE, where the non-reciprocal $I_c$ arises from a chemical potential shift induced by external line resistance. When a dc current flows through the circuit resistance, the resulting voltage drop shifts the chemical potential, creating an asymmetric $I_c$ without intrinsic symmetry breaking. This mechanism is particularly pronounced in Cooper-pair transistors, where $I_c$ is sensitive to Fermi level variations.[135] Furthermore, geometrical asymmetry combined with a thermal gradient in FeSe can produce a non-reciprocal $I_c$ through thermoelectric mechanisms, creating a diode behavior that is locked to the device geometry.[133] Theoretically, asymmetric charge accumulation at the Josephson capacitance also directly leads to a $B$-even zero-field SDE.[142]

### 3.3. Tunability
Non-reciprocal supercurrents in 2D superconductors exhibit high tunability under various external controls.[18, 143] Multiple tuning parameters including magnetic fields, electric fields, strain, device geometry, thermodynamic conditions, and microwave irradiation can effectively modulate the efficiency and polarity of SDE, as illustrated in Fig. 9. This broad tunability facilitates the investigation of fundamental mechanisms in superconducting systems. Additionally, it offers viable pathways for developing high-performance SC rectifiers and low-dissipation quantum electronic devices.

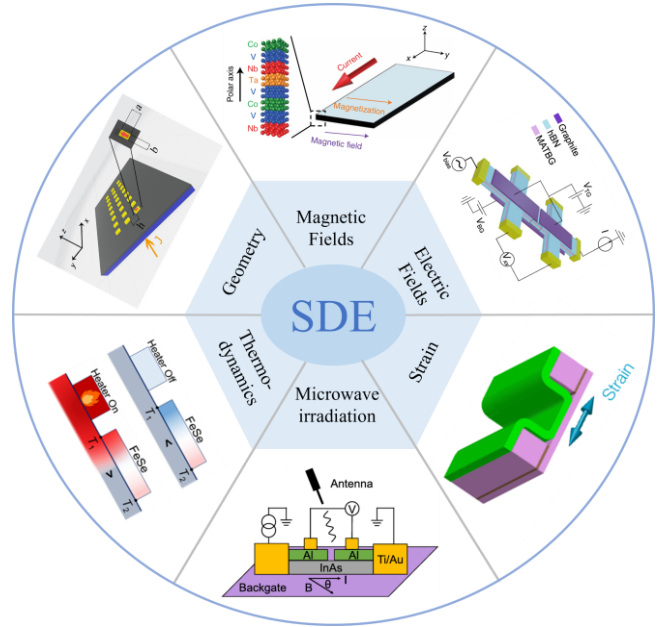

Fig. 9 Experimental tuning of the SDE via magnetic fields, electric fields, strain, device geometry, thermodynamic conditions, and microwave irradiation. Reproduced with permission from Ref. [86, 100, 111, 123, 133, 144]. Copyright 2025, the authors.

### 3.3.1. Magnetic fields

Magnetic fields serve as a versatile parameter for controlling the SDE. An in-plane $B$ primarily modifies the superconducting order parameter, generating FMCP (Fig. 4).[34-36] This mechanism allows for the continuous adjustment of diode efficiency and can produce polarity reversals at specific field values. Conversely, a perpendicular $B$ produces distinct vortex dynamics in type II superconductors. When combined with engineered asymmetric pinning potential, the perpendicular $B$ enables highly efficient diode operation with widely tunable rectification characteristics as realized in many systems (Fig. 6).[73, 86-88, 90-99]

Furthermore, magnetic fields provide a programming capability for non-volatile SDE. In superconducting magnetic heterostructures, external fields can configure the magnetization state of ferromagnetic layers to establish remanent diode effects that persist at zero field (Fig. 7). This approach has been demonstrated in [Nb/V/Co/V/Ta]$_{20}$ superlattices, kagome superconductors and MnBi$_2$Te$_4$-based JJs, facilitating switchable SDE polarity through magnetic history control.[100, 145, 146]

### 3.3.2. Electric fields

Electric fields control relies on the electrostatic modulation of fundamental SC

properties, such as SOC, $I_c$, and pair-breaking dynamics, which have been applied in various SC systems (e.g., cuprates, iron-based superconductors, and transition-metal dichalcogenides).[147-151] This capability enables the electric fields controlled SDE. Promising pathways have been demonstrated toward gate-controlled JJs,[50, 63, 152-158] as well as current-driven magnetization orientation.[40] These approaches collectively pave the way for programmable dissipationless electronics based on the SDE.

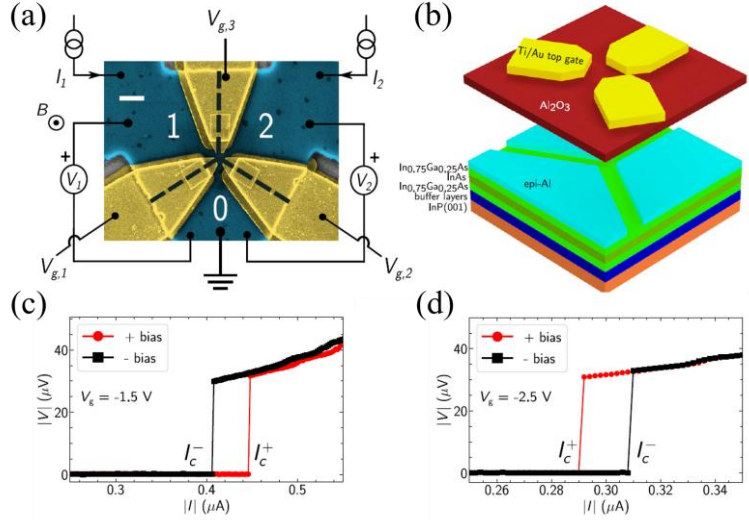

Fig. 10 (a) False-color SEM image of a three-terminal JJ with tunable gates (gold). The junction area between Al electrodes (blue) is outlined. (b) 3D schematic of the layered heterostructure. (c)(d) V-I curves at B = 0.04 mT, showing a reversal of the diode polarity as the gate voltage is tuned from -1.5 V (c) to -2.5 V (d). Reproduced with permission from Ref. [152]. Copyright 2023, the authors.

Based on three-terminal JJs, Gupta *et al*. demonstrate a highly gate-tunable SDE in an InAs two-dimensional electron gas proximitized by an epitaxial aluminum layer (Figs. 10(a) and 10(b)).[152] They achieve the full electrostatic control over both the efficiency and polarity of the SDE, realized by applying symmetric gate voltages to control the spatial distribution of the supercurrent density. This gate-driven reconfiguration effectively switches the diode polarity (Fig. 10(b)). In a different architecture, Nikodem *et al*. reported a gate-tunable SDE in a topological insulator nanowire-based nano-SQUID, where the diode efficiency and its sign are modulated by a back-gate voltage.[159] These works collectively highlight the pivotal role of electrostatic gating in realizing non-reciprocal supercurrent control.

### 3.3.3. Strain

Uniaxial strain serves as a powerful IS-breaking mechanism that enables SDE by reducing crystal symmetry and inducing electric polarization.[160] This is observed in the noncentrosymmetric trigonal superconductor PbTaSe$_2$.[111] While its crystal structure permits SHR, the SDE is absent in the unstrained state. However, when a uniaxial strain is applied along the zigzag direction with a parallel supercurrent, the SDE exhibits a *B*-odd symmetry (Figs. 11(a) and (b)). In contrast, when both strain and supercurrent are aligned along the armchair direction, the SDE displays *B*-even characteristics (Figs. 11(c) and (d)). The *B*-even behavior is particularly significant as it signifies a strain-induced polarization that induces the SDE without breaking TRS.

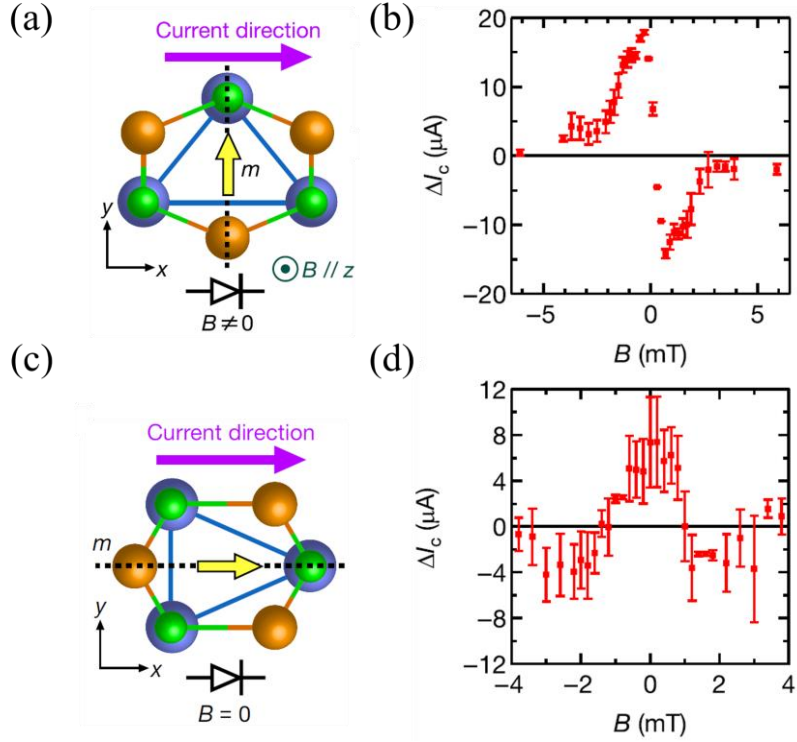

Fig. 11 (a) and (c) Schematic illustration of crystal structures of strained sample. Yellow arrows indicate the broken mirror plane due to trigonal symmetry. (b) and (d) *B* dependence of difference between $I_{c+}$ and $I_{c-}$ ($\Delta I_c = I_{c+} - I_{c-}$) in strained samples. Reproduced with permission from Ref. [111]. Copyright 2024, the authors.

The universality of strain-controlled SDE is further evidenced in centrosymmetric $NbSe_2$, where local strain engineering generates a robust zero-field SDE. Remarkably, strained $NbSe_2$ devices exhibit the *B*-even response consistently emerging for current along the armchair direction. These findings identify strain engineering as an effective method for controlling non-reciprocal superconductivity.[130]

### 3.3.4. Geometry

Geometric control represents a highly controllable and material-independent strategy for realizing the SDE.[73, 88, 90-92] Distinct from methods relying on intrinsic crystal asymmetry, this approach realizes non-reciprocal supercurrent transport by artificially breaking spatial inversion symmetry. For instance, conformal-mapped nanohole arrays, a scale-invariant pinning geometry, demonstrate strong rectification over broad temperature and field ranges, as shown by Lyu *et al*.[93] In such systems, the gradient distribution of pinning sites breaks global IS, guiding vortices preferentially in one direction and resulting in a giant tunable SDE (Fig. 12). Recently, using conductive atomic force microscope lithography, Wang *et al*. and Yu *et al*. demonstrated real-time, non-volatile editing of the SDE by nanoscopically altering the edge morphology of the SC channel.[161, 162]

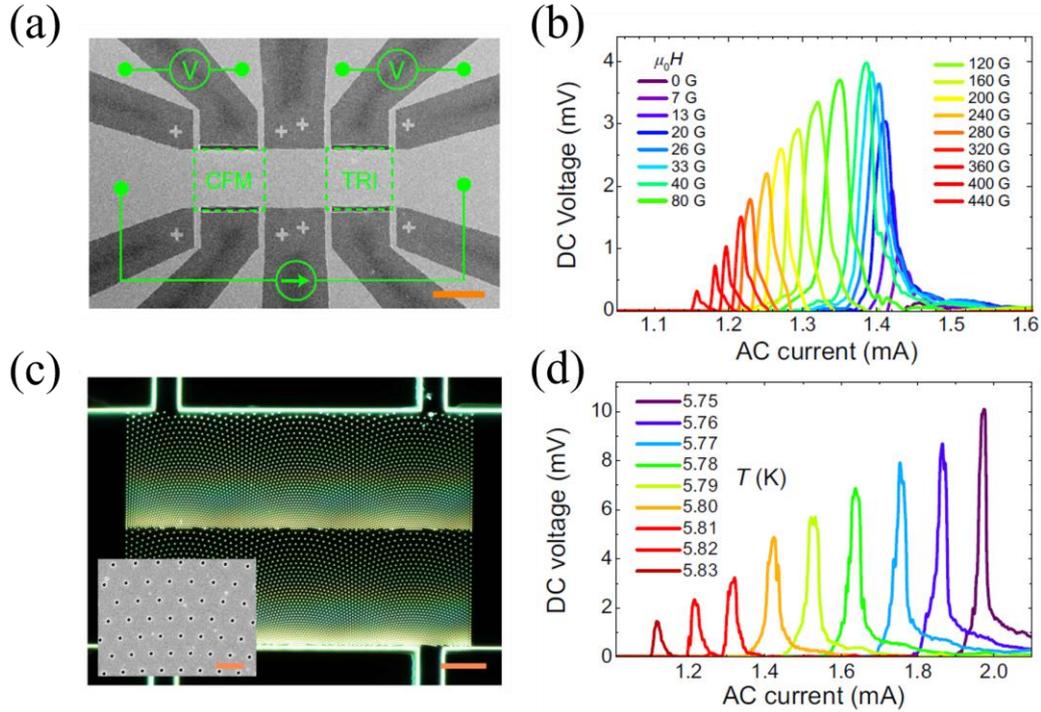

Fig. 12(a) Scanning electron microscopy image of a superconducting MoGe microbridge containing two sections with conformal mapped triangular and regular triangular arrays of nano-holes, respectively. The AC current is applied horizontally, and the DC voltage is measured with the top or the bottom leads. The magnetic field is applied perpendicularly to the sample plane. Scale bar, 40 $\mu$m. (b) Dark field optical image of the conformal arrays of nano-holes. Scale bar, 10 $\mu$m. The inset is a SEM image of the nano-holes with diameter of 110 nm. Scale bar of inset, 1 $\mu$m. (c) Voltage curves of the AC current dependence of the DC voltage at various magnetic fields. (d) Rectification amplitude at various temperatures in a magnetic field of 40 Oe. Reproduced with permission from Ref. [93]. Copyright 2021, the authors.

### 3.3.5. Thermodynamics

The thermodynamic behavior of SDE, particularly its response to thermal cycling and temperature gradients, plays a crucial role in revealing the nature of symmetry breaking and domain dynamics in the SC states. Recent studies across diverse material platforms, including kagome superconductors,[125] cuprates,[132] and iron-based superconductors,[133] have demonstrated that thermodynamics serves as an effective way for controlling the SDE, thus enabling thermal tunability of the diode polarity, efficiency, and operational stability.

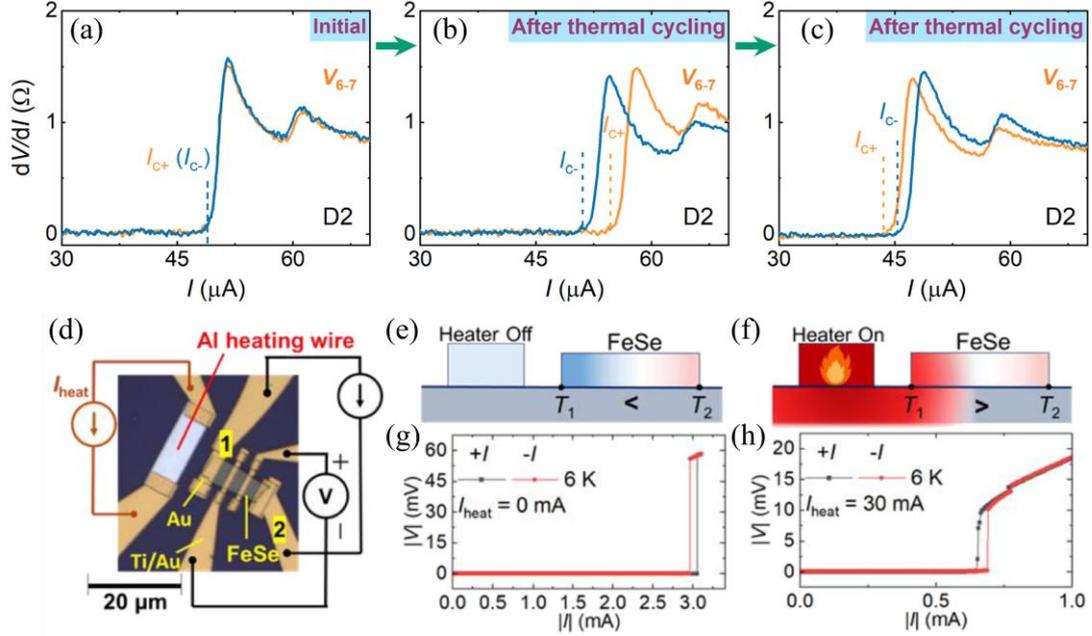

Fig. 13 (a)(b)(c) d$V$/d$I$ versus $I$ before and after thermal cycling. The device shows negligible non-reciprocity in the initial state. After thermal cycling, the SDE with stochastic polarity emerges. Reproduced with permission from Ref. [125]. Copyright 2024, the authors. (d) Optical image of the device and measurement configuration. (e) and (f) Schematic of the thermal gradient in FeSe when the heater is off (e) and on (f). (g) The SDE at 6 K without external heating, showing positive polarity ($I_{c+} > I_{c-}$) under the natural thermal gradient ($T_1 < T_2$). (h) SDE under applied heating (30 mA to Al wire), reversing the thermal profile to $T_1 > T_2$, and correspondingly inverting the diode polarity to negative ($I_{c+} < I_{c-}$). Reproduced with permission from Ref. [133]. Copyright 2025 American Physical Society.

In the kagome superconductor CsV$_3$Sb$_5$, thermal history directly governs the configuration of dynamic SC domains with spontaneously broken TRS.[125] The zero-field SDE exhibits distinctive polarity switching following thermal cycling across the $T_c$. These phenomena are attributed to the reorganization of chiral SC domains and the corresponding boundary supercurrents. Therefore, thermal protocols offer an effective thermodynamic approach for controlling the SDE in chiral superconductors (Figs. 13(a), 13(b) and 13(c)). A similar thermal-history dependence of SDE has also been observed in the cuprate superconductor BSCCO.[132]

In FeSe, a distinct SDE mechanism arises from the interplay of a large thermoelectric response and geometrical asymmetry (Figs. 13(d), 13(e), 13(f), 13(g) and 13(h)). Here, an in-plane temperature gradient ($\nabla T$) generated by Joule heating at asymmetric contacts induces an additional thermoelectric current, which non-reciprocally modifies the effective $I_c$. This effect is intrinsic and operates without external magnetic fields or magnetic layers. Furthermore, the SDE polarity in FeSe can be reversibly switched by manipulating the heat flow using an integrated micro-heater, which reverses the direction of $\nabla T$. This experiment provides direct evidence that the thermal gradient can effectively control the SDE.[133]

### 3.3.6. Microwave irradiation

Microwave irradiation offers a dynamic tuning method, capable of inducing non-equilibrium states through periodic AC driving to effectively modulate the directionality and efficiency of the SDE.[144, 163-165] A prominent demonstration was first provided in the work by Su *et al*.,[144] which studies Al-InAs nanowire-Al JJs under combined microwave irradiation and magnetic fields (Fig. 14). In the absence of microwave driving, the devices exhibit only a weak SDE. However, upon application of microwave irradiation, a pronounced horizontal offset of the zero-voltage step (the zeroth Shapiro step) in the *V-I* curves is observed. This offset current increases with microwave power and shifts the entire zero-voltage step away from the origin. At sufficiently high power, the system enters the unidirectional superconductivity regime, where supercurrent flows preferentially in one direction. These findings highlight microwave irradiation as a potent tool for enhancing the efficiency of the SDE.

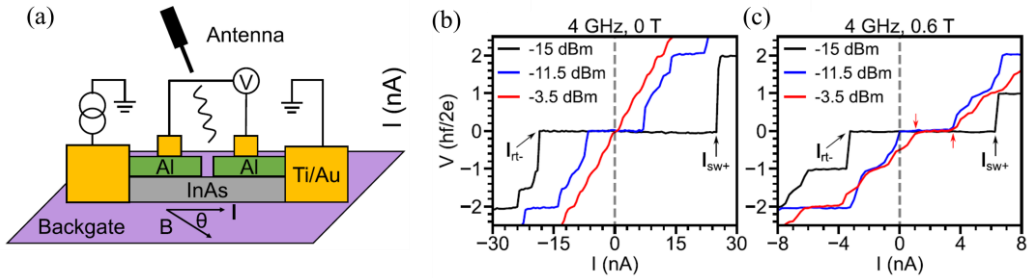

Fig. 14 (a) Measurement schematic with a fixed in-plane magnetic field angle $\theta$. (b) and (c) *V-I* curves under microwave irradiation at zero field (b) and finite field (c). Reproduced with permission from Ref. [144]. Copyright 2024 American Physical Society.

More recently, in the kagome superconductor $CsV_3Sb_5$ and BSCCO JJs, Lou *et al*. and Wang *et al*. also demonstrated a unidirectional SDE under microwave irradiation even at zero magnetic field. By exploiting the intrinsic Josephson diode effect and its response to microwave irradiation, the zeroth Shapiro step was entirely displaced to one side of the zero-current axis. According to the standard diode efficiency formula $\eta = (I_{c+} - |I_{c-}|)/(I_{c+} + |I_{c-}|)$, this configuration corresponds to 100% efficiency, representing an ideal superconducting rectifier that operates without any external magnetic fields.[163, 165]

### 3.4. Applications

Beyond its fundamental physical significance, the remarkable tunability of the SDE makes it a promising component for next-generation superconducting electronics. The capability to control both the efficiency and polarity of non-reciprocal supercurrents enables the design of innovative circuit architectures that outperform their semiconducting counterparts in energy efficiency and operational speed.

A representative application of the SDE involves high-efficiency rectifiers for alternating-current (ac) to direct-current (dc) conversion. Conventional semiconducting rectifiers are limited by resistive losses, whereas superconducting rectifiers can operate with negligible dissipation. Following early work demonstrated half-wave rectification in individual superconducting diode (SD) devices, recent progress has focused on realizing full-wave rectification using integrated bridge configurations, as shown in Figs. 15(a) and 15(c).[166-168] By arranging multiple SDs into a bridge, both half-cycles of an ac input can be rectified, significantly enhancing power conversion efficiency and output stability, as illustrated in Figs. 15(b) and 15(d). Such architectures are highly relevant for on-chip power management in cryogenic computing systems.

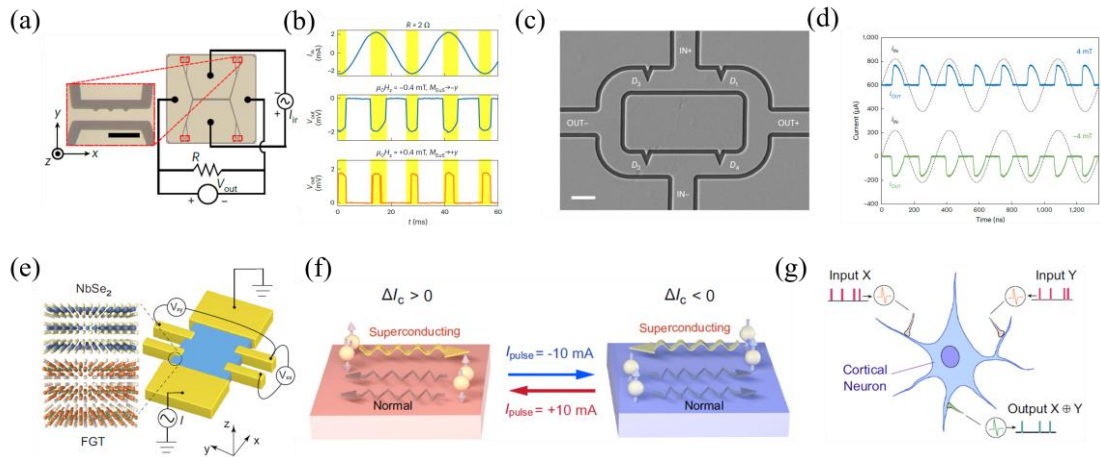

Fig. 15 (a) Left: an optical microscope image of the upper-right V/EuS superconducting diode (SD) within the rectifier circuit. The scale bar is 20 $\mu$m. Right: Schematic of the full superconducting rectifier. Red rectangles mark individual SDs, black lines indicate the rectifier current path, R denotes load resistance, and the outer frame size is 1 × 1 mm$^2$. (b) Experimental demonstration of reversible SD at 1.7 K. The upper panel shows time-dependent input current ($I_{in}$). The middle and bottom panels show the corresponding output voltage, $V_{out}$ at ±0.4 mT with EuS magnetization ($M_{EuS}$) aligned along ±y, respectively. Reproduced with permission from Ref. [166]. Copyright 2025 the Authors. (c) Scanning electron micrograph of a superconducting bridge rectifier fabricated on a 13-nm-thick NbN film. The NbN structures appear in grey, with dark borders outlining the underlying SiO$_2$ substrate. Scale bar: 1 $\mu$m. (d) Full-wave current rectification of a 3 MHz sinusoidal input ($i_{IN}$, black dashed traces) under both positive (blue) and negative (green) applied field. Reproduced with permission from Ref. [167]. Copyright 2025 the Authors. (e) Schematic of the SDE device composed of NbSe$_2$ and FGT flakes for electrical transport measurements. (f) Schematic of electrically switchable nonreciprocal superconducting transport, achieved after applying a current pulse. (g) Schematic structure of a biological cortical neuron, which performs classification of linearly nonseparable inputs via a nonlinear XOR operation. Reproduced with permission from Ref. [40]. Copyright 2024 the Authors.

Beyond analog signal processing, the non-volatile and electrically tunable characteristics of SDE offer novel pathways toward superconducting logic and neuromorphic computing. Recent studies on van der Waals SC heterostructures have demonstrated electrically switchable and non-volatile SDE polarity, which can be directly mapped to binary memory states or synaptic excitatory and inhibitory functions as depicted in Figs. 15(e), 15(f) and 15(g).[40, 104] Such capabilities form the foundation for non-reciprocal quantum neuronal transistors that mimic biological neurons. In such SD devices, input signals are integrated in a manner analogous to postsynaptic potentials. Upon exceeding a threshold, the SDE polarity switches, thereby controlling the propagation of supercurrent signals to subsequent stages. This biomimetic mechanism, operating with negligible dissipation, paves the way for ultra-low-power neuromorphic systems capable of complex cognitive tasks.

Moreover, incorporating tunable SDE into superconducting quantum interference devices (SQUIDs)[156, 169] and programmable JJ arrays[50, 170, 171] presents opportunities for intelligent sensors and programmable analog circuits. For instance, gate- or flux-tunable SDEs integrated within a SQUID can introduce directional control of magnetic flux, enhancing sensitivity and enabling novel signal processing paradigms.[172] The chirality inherent in non-reciprocal supercurrent transport further suggests potential applications in topological quantum circuits,[124-126, 173] where such directional behavior may facilitate the creation and manipulation of symmetry-protected quantum states.

## 4. Conclusion and perspectives

This review has systematically examined non-reciprocal transport in 2D superconductors, covering both dissipative and dissipationless SC regimes. We discussed the second-harmonic resistance (SHR), which arises from mechanisms including asymmetric vortex motion, spin-momentum locking, and asymmetric magnetic environments. For the dissipationless supercurrent diode effect (SDE), we first addressed field-induced mechanisms such as finite-momentum Cooper pairing, anomalous current-phase relations, and asymmetric vortex dynamics. We further explored zero-field SDE, classifying it into polarity-reversed types (governed by switchable magnetic order, unconventional superconductivity, and vortex pinning) and polarity-locked types (arising from p-n-type Josephson junctions, electric polarization, shift current, pinned moments, circuit-level effects, and asymmetric charge accumulation).

The SDE and related non-reciprocal phenomena provide a sensitive probe for investigating unconventional superconducting order parameters and hidden symmetries in quantum materials. Crucially, the SDE serves as a powerful diagnostic tool that directly reflects the microscopic physics of the system. By analyzing the sign, magnitude, and field dependence under various symmetry-breaking conditions, researchers can extract valuable information about key material properties. For instance, the evolution of diode efficiency with an in-plane magnetic field can map the strength of Rashba SOC.[39, 60, 174] The presence of a zero-field SDE can reveal the nature of spontaneous TRS breaking.[122-126] Moreover, the symmetry constraints of the diode response provide a stringent test for theoretical models of finite-momentum Cooper pairing,[43, 49] unconventional superconductivity,[125, 132] or topological phase.[159] Therefore, the SDE offers a unique transport-based window to uncover exotic quantum states, advancing the fundamental characterization of correlated quantum matter.

The highly tunable nature of the SDE, via magnetic and electric fields, strain, device geometry, thermodynamic conditions, and microwave irradiation, paves the way for designing low-dissipation superconducting electronic devices. These include superconducting rectifiers, logic circuits, and neuromorphic elements that could potentially replace their semiconducting counterparts in cryogenic applications. The development of gate-tunable and non-volatile SDE devices further highlights the potential for energy-efficient, high-speed superconducting electronics beyond conventional semiconductor technology. Future efforts should focus on integrating these functional elements into scalable quantum circuits, ultimately advancing both fundamental research and practical applications in dissipationless electronics.

## Acknowledgment

We are grateful to X. Lin and Q. Chen for helpful discussions. Project supported by National Natural Science Foundation of China (Grant No. 92565201) and the Fundamental Research Funds for the Central Universities.